%
%

\magnification=\magstep2

\hsize=6.46truein
\vsize=8.93truein
\hoffset=-0.43truein
\voffset=0.01truein

\pageno 1

\baselineskip=13.3pt

\def\gsim{\,$\raise0.3ex\hbox{$>$}\llap{\lower0.8ex\hbox{$\sim$}}$\,}
\def\lsim{\,$\raise0.3ex\hbox{$<$}\llap{\lower0.8ex\hbox{$\sim$}}$\,}
\def\vecS{{\vec S}}
\def\vecs{{\vec s}}

\noindent
\hbox{Version number: 1 (March 22, 1997)} \hbox{Paper~number:~10tonega}\break

\vskip -0.3truecm

\noindent
\centerline{\bf One-dimensional Heisenberg antiferromagnet in which}
\centerline{\bf two $S\!=\!1/2$ and two $S\!=\!1$ spins are arranged
alternatively}

\vskip 0.4truecm

\centerline{\bf T.~Tonegawa$^{\rm a}$, T.~Hikihara$^{\rm b}$,
T.~Nishino$^{\rm a}$,}

\centerline{\bf M.~Kaburagi$^{\rm c}$, S.~Miyashita$^{\rm d}$,
H.-J.~Mikeska$^{\rm e}$}

\noindent
$^{\rm a}$Department of Physics, Faculty of Science, Kobe University,
Rokkodai, Kobe 657, Japan

\noindent
$^{\rm b}$Division of Physics, Graduate School of Science and Technology,
Kobe University, Rokkodai, Kobe 657, Japan

\noindent
$^{\rm c}$Department of Informatics, Faculty of Cross-Cultural Studies,
Kobe University, Tsurukabuto, Kobe 657, Japan

\noindent
$^{\rm d}$Department of Earth and Space Science, Graduate School of Science,
\hbox{Osaka} University, Toyonaka, Osaka 560, Japan

\noindent
$^{\rm e}$Institut f{\"u}r Theoretische Physik, Universit{\"a}t Hannover,
30167 Hannover, Germany

\vskip 0.4truecm

\centerline{Abstract}

\parindent=1.5pc
Ground-state and thermodynamic properties of the one-dimensional Heisenberg
antiferromagnet in which two $S\!=\!1/2$ and two $S\!=\!1$ spins are arranged
alternatively are studied by a quantum Monte Carlo method and by analytical
estimates.  It is found that the ground state of the system undergoes a
second-order phase transition which accompanies the vanishing of the energy
gap.

\vskip 0.4truecm

\parindent=4.1pc
\item{Keywords:}

\parindent=1.0pc
\item{} Ground-state phase transition; Heisenberg antiferromagnet;
Mixed spin systems; One-dimensional systems; Quantum Monte Carlo simulation

\vskip 0.4truecm

\noindent
Address for Further Correspondence:

\parindent=1.0pc
\item{} Takashi Tonegawa

\itemitem{} Department of Physics, Faculty of Science, Kobe University,

\itemitem{} Rokkodai, Kobe 657, JAPAN.

\itemitem{} Tel.:~+81-78-803-0541$,\quad\quad$Fax:~+81-78-803-0522

\itemitem{} e-mail: tonegawa@kobe-u.ac.jp

\vfill\eject

\baselineskip=16.7pt

\parindent=1.5pc
One-dimensional quantum Heisenberg antiferromagnets have been the subject
of numerous studies over a long period, in particular since Haldane's
prediction [1,2] of the difference between the integer-spin and
half-integer-spin cases.  Almost all results of these studies support the
prediction, and now it is widely agreed that for isotropic nearest-neighbor
(nn) exchange, the former case is massive, while the latter case is
massless.  Under these circumstances, the $S\!=\!1/2$ and $S\!=\!1$
alternating spin chain with isotropic nn exchange, which is a simplest example
of mixed spin systems, has recently been of considerable interest both
experimentally [3] and theoretically [3-7].  A characteristic feature of this
chain is that it is a quantum ferrimagnet and its ground state is magnetic.

\parindent=1.5pc
In order to explore the mixed spin chain with a nonmagnetic ground state, we
discuss here the one-dimensional antiferromagnet in which two $S\!=\!1/2$ and
two $S\!=\!1$ spins are arranged alternatively.  We express the Hamiltonian
describing the system as
$$
    {\cal H} = \sum_{\ell=1}^{N/4}
          \bigl(J_1\,\vecs_{4\ell-3}\cdot\vecs_{4\ell-2}
              + J_2\,\vecs_{4\ell-2}\cdot\vecS_{4\ell-1}
              + J_3\,\vecS_{4\ell-1}\cdot\vecS_{4\ell}
              + J_2\,\vecS_{4\ell  }\cdot\vecs_{4\ell+1}\bigr)\,,  \eqno (1)
$$
where $\vecs_\ell$ ($\ell\!=\!1$, $2$, $5$, $6$, $\cdots$, $N\!-\!3$,
$N\!-\!2$) and $\vecS_\ell$ ($\ell\!=\!3$, $4$, $7$, $8$, $\cdots$, $N\!-\!1$,
$N$) are, respectively, the $S\!=\!1/2$ and $S\!=\!1$ spin operators, and $N$,
being assumed to be a multiple of four, is the total number of spins.  We
impose periodic boundary conditions ($\vecs_{N+1}\!\equiv\!\vecs_{1}$).  It
is noted that the ground state of the system is nonmagnetic except when
both $J_1$ and $J_3$ are negative.

\parindent=1.5pc
Applying the Wigner-Eckart theorem, we may conclude that in the limit of
$J_2\!\to\!\infty$ ($J_2\!\to\!-\infty$), the present system is equivalent to
the $S\!=\!1/2$ ($S\!=\!3/2$) bond-alternating chain with the exchange
constants ${1 \over 9}J_1$ and ${16 \over 9}J_3$ (${1 \over 9}J_1$ and
${4 \over 9}J_3$) as far as sufficiently low energy states are concerned.
Thus, the system is gapless when $J_2\!\to\!\infty$ and $J_1\!=\!16J_3\!>\!0$
and also when $J_2\!\to\!\infty$ and $J_1\!=\!4J_3\!>\!0$.  A perturbation
calculation taking into account the lowest singlet and triplet of each unit
cell ($\vecs_{4\ell-2}$, $\vecS_{4\ell-1}$, $\vecS_{4\ell}$,
$\vecs_{4\ell+1}$) leads to the result that, for sufficiently small
values of $\vert J_1 \vert$ and $\vert J_2 \vert$ with $J_3$ fixed at 1.0,
the energy gap vanishes
when $J_1\!=\!{4 \over 3}J_2^{\,2}\!+\!2 J_2^{\,3}\!+\!\cdots$.

\parindent=1.5pc
Confining ourselves to the case where \hbox{$J_1\!=\!J_3\!=\!1.0$} and
\hbox{$J_2\!>\!0$}, we investigated numerically the ground-state and
thermodynamic properties of the system by a quantum Monte Carlo (QMC)
method.  Let us first discuss the ground-state properties.  We carried out,
at sufficiently low temperatures ($T\!=\!0.05$ and $0.10$), a QMC
calculation without the global flip within the subspaces of
$M\!\equiv\!\sum_{\ell=1}^{(N/4)}\bigl(s_{4\ell-3}^z\!+\!s_{4\ell-2}^z\!+\!S_{4\ell-1}^z\!+\!S_{4\ell}^z\bigr)\!=\!0$ and $1$ to estimate the energies
$E(M)$ for $M\!=\!0$ and $1$, the energy gap $\Delta\!\equiv\!E(1)\!-\!E(0)$,
and the nn two-spin correlation functions
$\omega_{1,2}\!\equiv\!\langle s_1^z s_2^z\rangle$,
$\omega_{2,3}\!\equiv\!\langle s_2^z S_3^z\rangle$, and
$\omega_{3,4}\!\equiv\!\langle S_3^z S_4^z\rangle$, where
$\langle\cdots\rangle$ stands for the Monte Carlo average within the $M\!=\!0$
subspace.  In the calculation, $10^6$ Monte Carlo steps were performed after
$10^5$ initial steps for obtaining the thermal equilibrium.  The Trotter
number $n$ was chosen to be $n\!=\!12$, $16$, $24$, $32$, $40$, and $48$, and
the $n$-dependence of the QMC results was extrapolated to $n\!\to\!\infty$ by
making a least-squares fit to the
formula, $A_n\!=\!A_\infty\!+\!a_1/n^2\!+\!a_2/n^4$.  For both temperatures
of $T\!=\!0.05$ and $T\!=\!0.10$, almost the same extrapolated values were
obtained, and thus we consider that $T\!=\!0.05$ is low enough to discuss the
ground-state properties.  The QMC calculation was carried out for $N\!=\!8$,
$16$, $32$, $64$, and $128$, for various values of $J_2$.  As for $E(0)$,
$E(1)$, and $\Delta$ for $N\!=\!8$, we compared the extrapolated values for
$T\!=\!0.05$ with the values obtained by a method of exact diagonalization;
this comparison shows that the precision in the former values goes down to the
second decimal point.  In order to obtain the results in the thermodynamic
limit ($N\!\to\!\infty$), the extrapolated values for the latter four $N$'s
were further extrapolated by the least-squares method using a linear function
of $1/N^2$.

\parindent=1.5pc
Fig.~1 depicts the $J_2$-dependence of the energy
gap $\Delta$.  From this figure we see that there exists a massless point at
\hbox{$J_2\!=\!J_{2{\rm c}}\!\sim\!0.77$}, at which the ground state of the
system undergoes a second-order phase transition.  Giving a thought to the
$J_2$-dependences of the correlation functions $\omega_{1,2}$,
$\omega_{2,3}$, and $\omega_{3,4}$ depicted in Fig.~2, we may schematically
represent, by means of the valence-bond-solid (VBS) picture [8,9], the ground
states for $J_2\!<\!J_{2{\rm c}}$ and that for $J_2\!>\!J_{2{\rm c}}$ as shown
in Fig.~3.

\parindent=1.5pc
Next, we briefly discuss the thermodynamic properties.  Carrying out a QMC
calculation which takes the global flip into account, we calculated the
temperature dependence of the magnetic susceptibility and the specific
heat.  We obtained the result that the susceptibility per unit cell for
\hbox{$J_2\!=\!0.77$} takes a finite value in the $T\!\to\!0$ limit; this is
consistent with the fact that the energy gap vanishes at this value of
$J_2$.  We also found that for sufficiently large values of $J_2$ such as
\hbox{$J_2\!=\!5.0$} and $10.0$, the specific heat per unit cell versus
temperature curve has a double peak; the higher-temperature peak is associated
with the Schottky-type peak of the two-spin system of $S\!=\!1/2$ and
$S\!=\!1$ spins, while the lower-temperature one is associated with the
specific heat peak of an antiferromagnetic $S\!=\!1/2$ chain with bond
alternation.

\parindent=1.5pc
One of us (H.-J.~M.) gratefully acknowledges the support from
the Japan Society for the Promotion of Science.  The present work is supported
in part by a Grant-in-Aid for Scientific Research from the Ministry of
Education, Science, Sports and Culture, Japan.  We thank the Supercomputer
Center, Institute for Solid State Physics, University of Tokyo and the
Computer Center, Tohoku University for computational facilities.

\vfill\eject

\centerline{\bf References}

\parindent=1.3pc

\item{[1]} F.~D.~M.~Haldane:~Phys.~Lett.~{\bf 93A} (1983) 464.

\item{[2]} F.~D.~M.~Haldane:~Phys.~Rev.~Lett.~{\bf 50} (1983) 1153.

\item{[3]} O.~Kahn, Y.~Pei and Y.~Journaux, {\it Inorganic Materials},
ed.~D.~W.~Bruce and D.~O'Hare (John Wiley \& Sons, 1996) p.65.

\item{[4]} A.~K.~Kolezhuk, H.-J.~Mikeska and S.~Yamamoto:~Phys.~Rev.~B
{\bf 55} (1997) R3336.

\item{[5]} S.~Brehmer, H.-J.~Mikeska and S.~Yamamoto:~J.~Phys.:~Condens.~Matter
{\bf 9} (1997) 3921.

\item{[6]} S.~K.~Pati, S.~Ramasesha and D.~Sen:~Phys.~Rev.~B {\bf 55} (1997)
8894.

\item{[7]} F.~C.~Alcaraz and A.~L.~Malvezzi: preprint (cond-mat/9611227).

\item{[8]} I.~Affleck, T.~Kennedy, E.~H.~Lieb and
H.~Tasaki:~Phys.~Rev.~Lett.~{\bf 59} (1987) 799

\item{[9]} I.~Affleck, T.~Kennedy, E.~H.~Lieb and
H.~Tasaki:~Commun.~Math.~Phys. {\bf 115} (1988) 477.

\vfill\eject

\centerline{\bf Figure Captions}

{\leftskip=1.5pc
\parindent=-1.5pc
Fig.$\,$1.\ \ Plots of the $N\!\to\!\infty$ energy gap $\Delta$ versus
$J_2$.  The open circles show the QMC results, and the solid lines are guides
to the eye.  Note that $\Delta$ vanishes at
$J_2\!=\!J_{2{\rm c}}\!\sim\!0.77$.\par}

{\leftskip=1.5pc
\parindent=-1.5pc
Fig.$\,$2.\ \ Plots of the $N\!\to\!\infty$ correlation functions
$\omega_{1,2}$ (open circles), $\omega_{2,3}$ (open squares), and
$\omega_{3,4}$ (open diamonds), versus $J_2$, where each function is normalized
by its minimum value (note that $-1/4\!\le\!\omega_{1,2}\!\le\!1/12$,
$-1/3\!\le\!\omega_{2,3}\!\le\!1/6$, and
$-2/3\!\le\!\omega_{3,4}\!\le\!1/3$).  The solid lines are guides to the eye.
\par}

{\leftskip=1.5pc
\parindent=-1.5pc
Fig.$\,$3.\ \ Schematic representations of the ground states for (a)
$J_2\!<\!J_{2{\rm c}}$ and (b) $J_2\!>\!J_{2{\rm c}}$ by means of the VBS
picture.  The solid circles represent the $S\!=\!1/2$ spins, and two
$S\!=\!1/2$ spins connected by the solid line form a singlet pair.  Each open
ellipse surrounding two $S\!=\!1/2$ represents an operation of constructing
an $S\!=\!1$ spin from these $S\!=\!1/2$ spins by symmetrizing them.
\par}

\bye